# Augmented Reality in ICT for Minimum Knowledge Loss

Mr. RamKumar Lakshminarayanan
Department of IT,
HCT,
Muscat, Oman.

Dr. RD.Balaji
Department of IT,
HCT,
Muscat, Oman

Dr. Binod kumar
Department of IT,
HCT,
Muscat, Oman

Ms. Malathi Balaji
Department of IT,
HCT,
Muscat, Oman

*Abstract*—Informatics world digitizes the human beings, with the contribution made by all the industrial people. In the recent survey it is proved that people are not accustomed or they are not able to access the electronic devices to its extreme usage. Also people are more dependent to the technologies and their day-to-day activities are ruled by the same. In this paper we discuss on one of the advanced technology which will soon rule the world and make the people are more creative and at the same time hassle-free. This concept is introduced as 6$^{th}$ sense technology by an IIT, Mumbai student who is presently Ph.D., scholar in MIT, USA. Similar to this research there is one more research going on under the title Augmented Reality. This research makes a new association with the real world to digital world and allows us to share and manipulate the information directly with our mental thoughts. A college which implements state of the art technology for teaching and learning, Higher College of Technology, Muscat, (HCT) tries to identify the opportunities and limitations of implementing this augmented reality for teaching and learning. The research team of HCT, here, tries to give two scenarios in which augmented reality can fit in. Since this research is in the conceptual level we are trying to illustrate the history of this technology and how it can be adopted in the teaching environment.

*Keywords: Augmented Reality, 6$^{th}$ sense technology, Teaching and Learning, ICT*

## I. INTRODUCTION

Augmented Reality is a live, direct and indirect, view of a physical, real-world environment where the information about the surrounding real world of the user becomes interactive and digitally modified. [1]
Augmented Reality (AR) is taking digital or computer generated information, whether let it be images, audios, videos and touch or haptic sensations and overlaying them over in a real-time environment [2].
A. *Characteristics of Augmented Reality*
The three characteristics of augmented reality are as follows:
a. AR combines real and virtual information.
b. AR is interactive in real time.
c. AR operates and is used in a 3D environment.

## II. HISTORY OF AUGMENTED REALITY

In 1962, Morton Heilig, designed a multi-sensory technology that had visuals, sound, vibration and smell. It is a motorcycle simulator Sensorama.

A device paired to a headset such as harness or helmet is called head-mounted display (HMD). In 1968, Ivan Sutherland created an optical see-through HMD and one of the examples used six degrees-of-freedom. He called it as Sword of Damocles.

In 1975, Myron Krueger created Videoplace, which allowed users to interact with virtual objects. In 1992, Tom Caudell and David Mizell coined the term "Augmented Reality" at Boeing's Computer Services' Adaptive Neural Systems Research and Development project.

Markers are physical objects or places where the real and Virtual Environment are fused together. The idea of 2D matrix marker was developed by Jun Rekimoto in the year 1996. D' Fusion was created a product for Augmented Reality. 3D markers were presented by Mathias Mohring in Mobile phones in the year 2004. In the year 2006, Nokia initiated the image captured by the camera and annotated the users surrounding in real time with graphics and text.

Wikitude World Browser which combines the GPS and compass data with Wikipedia entries which overlays the information with smartphone camera was launched in the year 2008.

## III. AUGMENTED REALITY DEVICES

The main devices for Augmented Reality are displays, input devices, tracking and computers. The types of displays are head mounted displays (HMD), handheld displays and spatial displays. The types of input devices for AR systems are gloves, wireless wristband, smart phones with touch screen. The types of tracking devices are digital cameras, optical sensors, GPS, accelerometers, solid state compasses, wireless sensors etc., Earlier computers was used to process the camera images, but now with the advent of the smart-phone technology the usage of computers as back pack configuration is considerably reducing.

## IV. AUGMENTED REALITY INTERFACE

The interaction in AR applications is classified as tangible AR interfaces, collaborative AR interfaces, hybrid AR interfaces, and the emerging multimodal interfaces.

Direct interaction with the real world by exploiting the use of real, physical objects and tools is supported by Tangible interfaces. For the use of multiple displays to support remote and co-located activities collaborative AR interfaces are used. Hybrid interfaces combine an assortment of different, but complementary interfaces as well as the possibility to interact through a wide range of interaction devices. Multimodal AR Interfaces combine real objects input with naturally occurring forms of language and behaviors such as speech, touch, natural hand gestures, or gaze.

## V. AUGMENTED REALITY SYSTEMS

Fixed indoor systems, fixed outdoor systems, mobile indoor systems, mobile outdoor systems and mobile indoor and outdoor systems are the five categories of Augmented Reality Systems.

## VI. AUGMENTED REALITY MOBILE SYSTEMS

Augmented Reality Mobile Systems includes both the mobile phone applications and the wireless systems.

## VII. SIXTHSENSE / WUW - WEAR UR WORLD

*'SixthSense' is a wearable gestural interface that augments the physical world around us with digital information and lets us use natural hand gestures to interact with that information. By using a camera and a tiny projector mounted in a pendant like wearable device, 'SixthSense' sees what you see and visually augments any surfaces or objects we are interacting with. It projects information onto surfaces, walls, and physical objects around us, and lets us interact with the projected information through natural hand gestures, arm movements, or our interaction with the object itself. 'SixthSense' attempts to free information from its confines by seamlessly integrating it with reality, and thus making the entire world your computer. [4]*

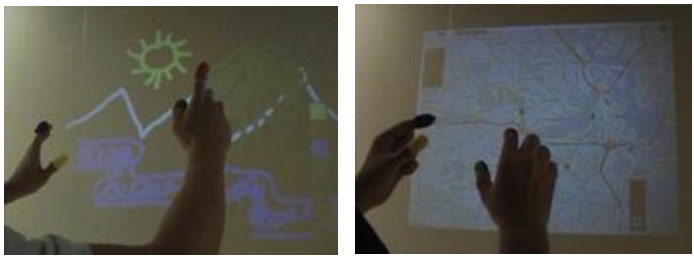

## VIII. OPTICAL CHARACTER RECOGNITION

Optical Character Recognition (OCR) is the mechanical or electronic translation of scanned images of handwritten, typewritten, or printed text, to machine encoded text. OCR is mainly used in language translation, digital libraries and even in the postal services. Now a day's most of the mobile phones are having high end camera functionality and means of enabling the features of OCR in mobile.

## IX. QR CODE

QR Code abbreviated from Quick Response Code was invented by Denso Wave, Japan. QR Code can detect the 2 Dimensional digital images. QR Code Reader can be a mobile phone to capture the dimensional images. The reader locates the three distinctive squares at the corners of the image, and uses a smaller square near the fourth corner to normalize the image for size, orientation and angle of viewing. The small dots are converted to binary numbers and their validity checked with an error-correcting code. QR Codes can be used with most of the mobile operating systems.

## X. SCENARIO 1

HCT is having 14,000 students studying in 8 departments in 8 buildings. Each department is having around 100 academicians and 20 other administrative staff. In administrative building there is more than 50 administrative staff members are working. Within each department there are at least 5 to 10 specializations are offered. Students and academicians are divided by their specialization. Again in each specialization students and staff are divided by levels like Diploma, Advanced Diploma and B.Tech., Every student is assigned with an advisor for allocating subject for the Academic semesters. Each course there will be a course coordinator and course teacher. Apart from that each student may take help from the administrative people of the department for their smooth progress in their studies. Many a times HCT is receiving complaints from students that they are facing problem in identifying the solution for solving some issues with a particular staff. Since staff members are in different staff room and cabin, it is difficult for a student to check where the particular staff seated or not. Similarly academic staff also complains that most of the time they spend in informing the students where a teacher is seated and the location of the building.

At this stage, we felt Augmented Reality can be applied to overcome the above problem particularly during the period of examination and time table registration of HCT. A student downloads the **HCT Identify Staff** app from the HCT website's mobile application page. By means of selecting student ID, student will be listed down his Advisory Name. The staff list, their specialization and QR Code with position information is available in the entrance board. The student will scan a QR code from the entrance board to find his staffs desks. This scenario highlights the potential of using QR codes for indoor AR navigation systems. Installation of a QR code is very low cost and easy to implement. Such codes can be installed in places where staffs change location so as to identify the staff's current location, while the student moves towards the staff desk he will be given direction by voice to match he reached the staff's location, it will be provided in AR view. AR view would be very intuitive so as to reduce navigation error and the time required for a student to understand the navigation information he is being informed.

## XI. SCENARIO 2

Any Technology will be successful only when it tempts or impresses a person to use it. Both in academic environment and administrative environment, this technology will give great impact when it is practiced for teaching and learning process. During the discussion with the research team, everyone felt that new technology should not be tried with the beginners as well as people at the exit level. Hence we have decided to take sample from the Advanced Diploma Level.

In the recent survey we have found students are facing problem in learning practical subjects like SQL concepts and Syntax. Here it is more difficult for the students to remember lot of syntax and commands. The research team decided to create Augmented Reality application which will automatically produce the SQL syntax when it scans data which needs to be stored in the database. For example: When a student scans a table structure as input with the mobile phone, the application should generate the corresponding SQL code as output. A sample of the table structure and SQL Code are given below:

A. *Proposed Table Structure:*

<u>**Student_Mark**</u>

| Column Name | Data type | Size | Constraint |
|---|---|---|---|
| Stud_id | Number | 9 | Primary key |
| Stud_Name | Varchar2 | 25 | |
| Prog_id | Varchar2 | 10 | Unique |
| Course_id | Varchar2 | 8 | |
| Quiz1 | Number | 5,3 | |
| Mid Exam | Number | 5,3 | |
| Final | Number | 5,3 | |
| Total | Number | 6,3 | Check <100 |
| Grade | Varchar2 | 2 | |
| Result | Varchar2 | 10 | Check "pass" or "fail" |

B. *Expected Code:*

Create table Student_Mark (
Stud_id number(9) primary key,
Stud_Name Varchar2(25),
Prog_id Varchar2(10) Unique,
Course_id Varchar2(8),
Quiz1 Number(5,3),
Mid_Exam Number(5,3),
Final Number(5,3),
Total Number(6,3) constraint SMCH1 check total < 100,
Grade Varchar2(2),
Result Varchar2(10) check SMCH2 check (result = 'pass' or 'fail'));

The proposed system will be using OCR capture technology. The text is printed on the paper with a specific format in the fixed height and width captured by OCR. The application will capture and rectify images will be fed in to the OCR Engine. This application will use the mobile device's camera to capture the images (like smart phone camera). Once the OCR process is over, the syntax engine will collect the information from the process image and create the SQL query.

XII. CONCLUSION

This paper mainly concentrates on the Augmented Reality and the 6$^{th}$ sense technology due to the advantages of simplicity in this technology. This technology can be implemented in the near future with the minimum requirements of the resources, compared to the 6$^{th}$ sense technology. Still we felt it is not justified if we leave 6$^{th}$ sense technology without mentioning here. In the Augmented Reality we have mentioned the history, devices and interfaces. HCT Research reveals that 95% of the students are using their smart phones or mobile devices for their day-to-day learning process. HCT is also encouraging students to use E and M learning devices. The Augmented Reality device section gives confidence to us about the implementation of this technology. Most of the features required by the Augmented Reality are there with the smart phones in the recent days. The scenario's specified here are just a conceptual proposal by the research team of HCT, to successfully implement this new technology and to evaluate the improvements in the teaching and learning process. The next stage is to evaluate the knowledge loss in the learning process by this technology. It is obvious that any new technology may have some negative impacts in future that also to be evaluated after the implementation of this new technology.

REFERENCES

[1] Wikipedia.org accessed on Dec 30, 2012
[2] Greg Kipper, Joseph Rampolla, Augmented Reality: An Emerging Technologies Guide to AR, Elsevier, Dec 27, 2012
[3] Borko Furht, Handbook of Augmented Reality, Springer, Jan 1, 2011
[4] http://www.pranavmistry.com/ accessed on Dec 31, 2012
[5] Sonia Bhaskar et al., Implementing Optical Character Recognition on the Android Operating System for Business Cards, "EE 368 Digital Image Processing Notes" Spring 2010
[6] B. Girod. "EE 368 Digital Image Processing Notes," EE 368 Digital Image Processing Spring 2010.
[7] Gee Andrew et al., A topometric system for wide area augmented reality. Computers and Graphics 2011
[8] P. Serrano-Alvarado, C. Roncancio and M. Adiba, "A Survey of Mobile Transactions," Distributed and Parallel Databases, September 2004
[9] Fröhlich P, Oulasvirta A, Baldauf M, Nurminen A. "On the move, wirelessly connected to the world", Commun ACM 2011
[10] Henrysson A, Ollila M, Billinghurst M. "Mobile phone based AR scene Assembly". In: Proc 4th Int Conf Mob Ubiquitous Multimedia - MUM '05, ACM Press; 2005
[11] Reilly DF, Inkpen KM, Watters CR. "Getting the Picture: Examining How Feedback and Layout Impact Mobile Device Interaction with Maps on Physical Media", In: Int Symp Wearable Comput - ISWC '09, IEEE Press; 2009
[12] Costabile M, Angeli AD, "Explore! possibilities and challenges of mobile learning", In: Proc 26th Annu Int Conf Hum Comput Syst. – CHI '08, ACM Press: 2008

[13] http://www.rummble.com
[14] Spohrer J., Information in Places, IBM System Journal 38(4), 1999
[15] Acquisti, A.and Gross, R., Imagined Communities : Awareness, Information Sharing and Privacy on the Facebook. PET 2006.
[16] Gogging G., Cell Phone culture: Mobile technology in everyday life, Routledge, New York 2006.
[17] Greene K., Hyperlinking reality via phones. MIT Technology Review 2006.
[18] L. Bonanni, M. Seracini, X. Xiao, M. Hockenberry, B.C. Costanzo, A. Shum, R. Teil, A. Speranza and H. Ishii, International Journal of Creative Interfaces and Computer Graphics, 2010
[19] D.M. Popovici, R. Querrec, C.M. Bogdan and N. Popovici, International Journal of Computers, Communications & Control, 2010
[20] Langlotz Tohias, Degendorfer Claus, Mullone Alessandro, Schall Gerhard, Reitmayr Geehard, Schmalstieg Dieter, Robust Detection and tracking of annotations for outdoor augmented reality browsing, Computer and Graphics 2011.
[21] Philbin J., Chum O., Isard M., Sivic J., and Zisserman A., Object retrieval with large vocabularies and fast spatial matching. In Proc of CVPR, 2007.
[22] Phibin J., Chum O., Isard M., Sivic J., and Zisserman A., Lost in quantization: Improving particular object retrieval in large scale image databases. In Proc of CVPR, 2008.
[23] Swan J.E and Gabbard J.L., Survey of User-Based Experimentation in Augmented Reality, presented at Ist International Conference on Virtual Reality, Las Vegas, Nevada, 2005.
[24] Sivic J and Zisserman A., Video google: A text retrieval approach to object matching in videos. In Proc of ICCV, 2003.
[25] Wagner D., Langlotz T. and Schmalstieg D., Robust and Unobstructive marker tracking on mobile phones. In Proc. Of ISMAR'08, 2008.

## AUTHORS PROFILE

**Ramkumar Lakshminarayan**: He is post grauduate in Computer Science and at present working as a Lecturer, Computer Science in Higher College of Technology, Muscat. He is having 14 years of experience in teaching, consulting and software development. He has conducted training for leading corporate companies in India and abroad in the field of Database, Datawarehousing, Cloud Computing and Mobile Technology. He has presented articles in various journals around the Globe. He has did research in the field of Applications of Computers Science in the Management of AAVIN Dairy Cooperatives and submitted thesis to Bharathidasan University, India. He has conducted workshop in events of Free and Open Source Software.

**Malathi Balaji:** She did her Master of Computer Science from Anna University with Gold medal. Having rich experience in teaching at graduate and post graduate level for more than 8 years. Presently waiting for Ph.D., registration with reputed university. She has published many papers in national and international journals during her studies. She has proved her excellence in education from her childhood by scoring district ranking in 10$^{th}$ and 12$^{th}$, as well as distinction in her UG. Presently doing research in Networks field and certified by CISCO. She has worked in abroad also as corporate trainer.

**Dr. RD.Balaji:** He has completed his Ph.D., in the year 2010 in Computer Science. Preceding to this PhD., completed his Bachelors and Masters degree from Madurai Kamaraj University. He is having totally fifteen years of teaching at UG and PG level including ten years of abroad experience. Presently working in Higher College of Technology, one of the prestigious Colleges in the Sultanate of Oman. Published many papers in national and international Journals. He has visited more than 8 countries to present his research work. He has guided many M.Phil., students to do their research. He is in the process of getting guideship from universities. He evaluated many Ph.D., thesis as a foreign examiner. Having membership with more than ten international computer oriented institutions and member of editorial board and reviewer for many journals and conferences.